\input{aipcheck}

\documentclass[
  ,draft            
  ]
  {aipproc}

\layoutstyle{6x9}

\begin{document}

\title{Neutron Stars in Supernovae and Their Remnants}


\author{Roger A. Chevalier}{
  address={Department of Astronomy, University of Virginia, P.O. Box 400325, Charlottesville,
VA 22904, USA}
}

\begin{abstract}
 The magnetic fields of neutron stars have a large range ($\sim 3\times 10^{10}- 10^{15}$ G).
There may be a tendency for more highly magnetized neutron stars to come from more massive stellar progenitors, but other factors must also play a role.
When combined with the likely initial periods of neutron stars, the magnetic fields imply a  spindown power that covers a large range and is typically dominated by other power sources in supernovae.
Distinctive features of power input from pulsar spindown are the time dependence of power and the creation of a low density bubble in the interior of the supernova; line profiles in the late phases are not centrally peaked after significant pulsar rotational energy has been deposited.
Clear evidence for pulsar power in objects $<300$ years old is lacking, which can be attributed to large typical pulsar rotation periods at birth.

\end{abstract}
\maketitle


\section{Introduction}

In the late 1960's, pulsars were discovered in the Crab Nebula and the Vela supernova remnant, proving the relationship between neutron star formation and supernovae.
These remained the only such associations for many years, but now, especially as a result of X-ray observations, there are many more.
There have also been discoveries extending the known range of neutron star and  supernova properties.
These developments allow a fresh look at the role of neutron stars in supernovae and their remnants.

\section{The Variety of Neutron Stars}

The effect of a neutron star (NS) on its surrounding supernova and the observability of the neutron star depend on the magnetic field strength and the rotation period of the neutron star.
Recent studies of the radio pulsar population find a normal distribution of initial rotation period with $<P_0>\sim 300$ ms and $\sigma_{P_0}\sim 150$ ms and   a lognormal distribution of magnetic field with $<\log(B_0/G)>\sim 12.65$ and $\sigma_{\log(B_0)}\sim 0.55$ \cite{fkaspi06}.
When magnetars are included in the neutron star population, the range of inferred magnetic field for the population, including thermally emitting isolated NSs, radio pulsars, and magnetars, is increased.
Popov et al. \cite{popov10} find a lognormal distribution with $<\log(B_0/G)>\sim 13.25$ and $\sigma_{\log(B_0)}\sim 0.6$; about 10\% of NSs are born as magnetars in their model.
The highly magnetized neutron stars are not generally radio emitters, although there is some overlap between the two populations.

Another approach to the population of NSs is to examine the compact sources at the centers of young ($\le 3000$ yr old) core collapse supernova remnants.
The types of compact objects and examples of each are:
\begin{itemize}
\item
Rotation powered pulsars (or compact sources with wind  nebulae) with external interaction: G21.5--0.9, 0540--69, MSH 15-52, Kes 75, G11.2--0.3
\item
Rotation powered pulsars (or compact sources with wind  nebulae) without external interaction: Crab Nebula, 3C58, G54.1+0.3
\item
Magnetars: Kes 73, CTB 37B, G327.24+0.13
\item
Compact central objects: Cas A, Puppis A, G350.1--0.3
\item
No compact object detected: 1E 1202.2--7219.
\end{itemize}
Most of these objects are described in \cite{chev05}, where references can be found.  
More recent discoveries involve the neutron stars in G21.5--0.9 \cite{gupta05,camilo06}, CTB 37B \cite{HG10b},
G327.24-0.3 \cite{GG07}, and G350.1--0.3 \cite{gaens08}.
The rotation powered pulsars (RPPs) have been divided into 2 groups depending on whether the interaction with the external medium has been detected; deeper observations may result in the observation of such interaction, as in the case of G21--0.9 \cite{math05}.
Among the rotation powered pulsars studied in \cite{chev05}, the  range of estimated initial periods is $10-100$ ms, with an average of 40 ms.
These estimates are smaller than the estimate of the mean from the radio pulsar population \cite{fkaspi06},
suggesting that only the high $\dot E$ part of the young pulsar distribution is being observed.
Young pulsars with low $\dot E$ and weak external interaction would be difficult to detect.

An important recent development is the study of X-ray periods and $\dot P$'s for 3 compact central objects (CCOs).
The results on the period and magnetic field are 105 ms and $3.1\times 10^{10}$ G for Kes 79 \cite{HG10a}, 112 ms and $<9.8\times 10^{11}$ G for Puppis A \cite{GH09}, and 424 ms and $<3.3\times 10^{11}$ G for G296.5+10.0 \cite{GH07} .
Pulsations have not been detected from the compact X-ray source in Cas A, but a recent interpretation of the X-ray spectrum in terms of a carbon atmosphere indicates a magnetic field $<8\times 10^{10}$ G due to the lack of line features \cite{HH09}.
These results imply that there is a significant population of low magnetic field neutron stars that are not included in the radio pulsar population because of their low radio luminosities.
The numbers are small, so it is not possible to draw firm conclusions about the distribution of pulsar magnetic fields.
As with pulsars with long initial periods, central compact objects in remnants with weak external interaction are probably missing from the current observed population.
The 3 observed CCOs with periods have such low magnetic fields that little evolution in period is expected over the age of the remnant.
The 3 objects have an average $P$ of 214 ms, which is close to the mean of 300 ms estimated for young radio pulsars \cite{fkaspi06}.
In this case, the X-ray luminosity is thermal and may not be related to the spin rate, so that the periods are more representative than for RPPs.

Magnetars are observed to have long periods $\sim 5-12$ s and are presumed to be strongly spun down, so there is little information about their initial periods in their current spin parameters.
In the dynamo theory for the buildup of the high magnetic field, the initial period is $1-2$ ms and the rotational energy is rapidly deposited in the supernova \cite{DT92}.
The prediction of this model is that the supernovae should be of high energy and magnetars could have high space velocities \cite{DT92}.
These predictions have not been confirmed \cite{vink06}, so a ms birth period does not appear to be required for the formation of a magnetar.

For magnetic dipole radiation with a neutron star radius of $10^6$ cm and $\sin^2{\alpha}=1/2$, where $\alpha$ is the angle between the magnetic and the rotation axes,  the spindown power of a pulsar is $\dot E\approx 6\times 10^{34}B_{13}^2(P/300~{\rm ms})^{-4}$ ergs s$^{-1}$, where $B_{13}$ is the pulsar magnetic field in units of $10^{13}$ G and $P$ is scaled to the mean found in \cite{fkaspi06}.
For $B=10^{11}$ G and $P=100$ ms, $\dot E=5\times 10^{32}$ ergs s$^{-1}$, which is less that the typical thermal luminosity emitted by CCOs.
Thus, such a neutron star is plausible for the remnant 1E 0102.2 in the SMC, which is found to have an X-ray luminosity that is below that of the Cas A CCO \cite{rutkowski10}.
Kaplan et al. \cite{kaplan04} have set limits on central sources in shell remnants below 1/10 the Cas A CCO luminosity.
In the case of 1E 0102.2, a massive star event is indicated for the young remnant, whereas Type Ia events or a central black hole remain  possibilities for the others.
However, the likely low $\dot E$s for pulsars can easily account for faint central sources.

\section{Progenitors and Magnetic Fields}

The results described in the previous section show that the range of magnetic fields for young neutrons is large: from $3\times 10^{10}$ G to $~10^{15}$ G.
The factors that determine the range are not understood, but the mass of the progenitor star may be related to the field strength.
An indication of this came from the finding that the magnetars CXO J164710.2-455216 and SGR 1806-20 are associated with young stellar clusters that imply a progenitor mass $>(40-50)~M_{\odot}$ \cite{muno06,figer05}.
However, the magnetar SGR1900+14 was found in a cluster with lower mass stars, implying a progenitor mass of $17\pm 2~M_{\odot}$ \cite{davies09}.
A magnetar lies within  the young remnant Kes 73 \cite{VG97}, which shows evidence for circumstellar interaction that can be interpreted as indicating a Type IIL/b progenitor
\cite{chev05}; this would imply an intermediate mass progenitor.

The technique of using an associated cluster to estimate progenitor mass was recently used to estimate the progenitor mass of G54.1+0.3, yielding a mass of $\sim 17~M_{\odot}$ (B. Koo, in prep.); in this case, the $P$ and $\dot P$ of the pulsar imply a magnetic field of $1\times 10^{13}$ G.
In other cases associated clusters are generally not observed, so that other means must be used to estimate the progenitor  mass.
There are number of ways in which the core collapse supernova Type can be inferred from observations of young supernova remnants \cite{chev05}.  The main Types are IIP, where the explosion occurs with most of its H envelope intact, IIL/b in which only a fraction of the H envelope is left at the time of the explosion, and Ib/c in which the H envelope is completely lost.
For single stars, this listing of Types is in order of increasing progenitor mass, but the later Types may occur at lower masses for stars in close binary systems.

The compilation of objects in \cite{chev05} did not show a clear trend of magnetic field strength with Type.
However, 0540--69 was listed as a Type Ib/c event, but recent observations show that it has low velocity hydrogen \cite{sera05}, so it is more likely to be a Type IIP.
With this change, there does appear to be a trend with Type, such that the likely more massive progenitors leave more strongly magnetized pulsars.

Other recent observations have led to new information in this area.
As mentioned in the previous section, the X-ray spectrum of the Cas A central source suggests a low magnetic field, $<8\times 10^{10}$ G.
Observations of the light echo of the Cas A supernova have shown that it was a Type IIb event, like SN 1993J \cite{krause08}; the supernova had been previously inferred to be of this Type
from the properties of the supernova remnant \cite{chev05}.
Nucleosynthesis constraints suggest a progenitor mass of Cas A of $(15-25)~M_{\odot}$ \cite{young06}.
The Puppis A remnant, which also has a neutron star with low magnetic field, was also inferred to have a Type IIL/b supernova based on the circumstellar interaction and the composition of freely expanding ejecta \cite{chev05}.
On the other hand, the Crab Nebula, which has a pulsar with an estimated field of $4\times 10^{12}$ G, is inferred to have a progenitor mass of $(8-10)~M_{\odot}$ from nucleosynthesis arguments.
The progenitor mass is lower that that of the CCO's, but the magnetic field is higher.
Over the entire range of magnetic field ($3\times 10^{10}-10^{15}$ G), there is some tendency
for more massive progenitors to yield a more highly magnetized neutron star, but it is clear
that other parameters, such as rotation, metallicity, or binarity, must also play a role.

\section{Neutron Stars in Supernovae}

In SN 1987A, neutrinos were observed over a time of 10 s, indicating the presence of a neutron star on that timescale \cite{arnett89}.
The youngest neutron star in a remnant is the Cas A compact object, so there is period from $10-10^{10}$ s over which neutron stars are not clearly observed.
Although direct evidence is lacking, there is the expectation that pulsar power should be present in supernovae and so there has been attention to the ways in which that power might manifest itself.

The notion that pulsar power is responsible for the light curves of core collapse supernova was proposed soon after the discovery of pulsars.
Initial models attributed the explosion energy as well as the light to the pulsar \cite{bo74}, but the sweeping of all the progenitor mass into a shell, as expected in this case, is not compatible with observations.
Subsequent models assumed that there was an initial explosion and that the pulsar power was responsible for the light curve \cite{gaffet77a,gaffet77b}.
However, models with an instantaneous explosion in a massive star at the end of its evolution with allowance for power from radioactivity have been widely successful in reproducing supernova light curves and spectra, so that pulsar models have not been favored.

The discovery of peculiar and luminous supernovae has brought back the idea of pulsar power \cite{maeda07,kasen10,woosley10}.
With  magnetar power, the timescale for power input from the pulsar may be comparable to the diffusion time for radiation, so that the pulsar rotational energy can be efficiently turned into radiation for the light curve.
Some assumptions are made in these models.
One is that the spindown power for the pulsar is given by the formula for a rotating magnetic dipole in a vacuum.
A newly formed neutron star would find itself in an especially dense environment.
Fallback of matter from the surrounding supernova can accrete to the neutron star; this effect may be important for pulsar magnetic fields $\sim 10^{12}$ G, but at magnetar fields, the magnetic effects are likely to dominate near the neutron star \cite{chev89}.
The application of the spindown formula still requires that supersonic/superAlfvenic flow develop around the pulsar so that the vacuum solution applies.
In the standard MHD picture for pulsar nebulae, the flow in the wind nebula must decelerate from mildly relativistic velocities near the wind termination shock to the velocity of the outer boundary.
For the case of a slow outer boundary, as expected for a recently formed pulsar, this is likely to require a pulsar wind with relatively small magnetization.

Observed pulsar wind nebulae are not especially radiatively efficient.
Most of the pulsar spindown power goes into the internal energy of the shocked wind bubble and the kinetic energy of the swept up shell \cite{chev05}.
The nebula with the highest radiative efficiency is the Crab Nebula, which radiates $\sim 1/3$ of the pulsar spindown power in synchrotron radiation; the high efficiency is due to the high magnetic field resulting from the high spindown power and youth of the system.
For the very young pulsars considered here, synchrotron radiation losses are expected to be even more efficient.
Inverse Compton losses of energetic particles in the supernova radiation field are also efficient at early times.

Another issue is the degree of asymmetry of the pulsar nebula.
Models of gamma-ray bursts with magnetar power have shown that relativistic jets can be produced for a $10^{15}$ G neutron star with a $(1-2)$ ms initial rotation period \cite{kom07,bucc08}.
An important effect in producing the collimation is the buildup of toroidal magnetic flux in the shocked wind nebula and the hoop stresses of the toroidal flux.
Once the jet breaks out of the star, most of the pulsar power goes into jets and not into the supernova envelope.
The models considered by \cite{kasen10} have initial pulsars with lower magnetic fields, but comparable rotation energies and asymmetries could limit the transfer of power to the supernova envelope.

Given the parameters ejecta mass $M$ and pulsar initial spin $P_0$ and magnetic field $B$,  values can be found that approximately reproduce the peak luminosity and timescale of very luminous supernovae \cite{kasen10}.
In particular, the values $M=5~M_{\odot}$, $P_0=2$ ms, and $B=2\times 10^{14}$ G give a reasonable fit to the luminosity of SN 2008es \cite{kasen10,miller09,gezari09}.
However, the luminosity does not provide a strong constraint on the model.
Another source of power, such as interaction with dense mass loss, could presumably give a similar result if the mass loss parameters could be varied.
What is needed are properties of the magnetar model that are characteristic of this mechanism.
With the assumption of evolution with constant braking index $n$, the pulsar power
has the time dependence
$\dot E \propto (1+t/t_p)^{-(n+1)/(n-1)}$,
where $t_p$ is the initial spindown timescale.
Measured braking indices cover a range from 1.4 to 2.9 (see \cite{chev05} for references).
The light curve thus evolves to a power law with time in the declining phase, with the power law index ranging from 2 to 6, if the braking indices from pulsars with ages $10^3-10^4$ yr are applicable to the early times considered here.
As discussed by \cite{gezari09}, the decline of SN 2008es is approximately exponential and  drops below the magnetar model at an age $\sim 100$ days \cite{kasen10}.
Another supernova, SN 2010gx, had similar luminosity, light curve shape, and temperature to SN 2008es, and was followed to a fainter magnitude relative to the peak \cite{past10}.
The sharp drop of 5 magnitudes  
appears to be incompatible with the slow  decline expected with power input from $^{56}$Co decays or spindown of a pulsar, even with $n=1.4$.
The light curve can presumably be reproduced by interaction with dense mass loss near
the supernova, although the required density distribution is ad hoc.
One difference between SN 2010gx and SN 2008es is that SN 2010gx did not show lines of H or He.
However, SN 2010gx is representative of a group of events \cite{past10} and their similar properties suggests that the same mechanism may be operating in all of them.

Other distinguishing properties of pulsar power may come from spectroscopic observations.
With pulsar power in a supernova, the power goes into inflating a wind bubble in the supernova interior, bounded by an accelerating shell \cite{chev77}.
The shell acceleration stops when most of the pulsar rotational energy has been deposited so that the shell freely expands with the supernova ejecta.
Using their magnetar model, Kasen and Bildsten \cite{kasen10}  modeled the luminosity evolution of SN 2007bi, which was suggested to be a pair instability supernova with power input from radioactivity \cite{galyam09}.
The light curve showed a decline over 100's of days that was consistent with radioactivity, but the late emission expected from pulsar power also gives a reasonable fit.
A  distinguishing feature is provided by the late spectroscopy, which show centrally peaked lines, including Fe lines, as expected  for radioactive power  input.
In the pulsar power case, there should be relatively little material at low velocity.
Even though pair instability supernovae had not been expected for stripped envelope stars, which is the case for SN 2007bi, radioactivity is the preferred mechanism.
This is also the case for SN 1998bw, which \cite{woosley10} investigated as a possible supernova light curve with pulsar power, but noted that radioactivity provided a more natural explanation for the luminosity evolution.
The late spectra of SN 1998bw show centrally peaked line emission \cite{patat01}, which is another problem for a pulsar interpretation.

Even if pulsar power does not dominate the supernova light, there is the possibility that it plays some  role, especially at late times when the main supernova light has faded.
Strong limits have been set in the case of SN 1987A, where the lack of optical evidence for a compact object suggests a power $<8\times 10^{33}$ ergs s$^{-1}$ \cite{graves05}; searches for a compact object at radio and X-ray wavelengths have also come up empty \cite{ng09}.  
As the ejecta expand, there is a better chance to observe through the ejecta to the central compact object, but the strengthening interaction with the circumstellar medium makes observations of the center more difficult.
Continuing circumstellar interaction in supernovae generally hinders the observation of a central compact object.
SN 1986J showed signs of a central compact object at radio wavelengths, but recent observations show that circumstellar interaction is a possibility \cite{bieten10}.
Perna et al. \cite{perna08} have recently gathered X-ray observations of supernovae; most are upper limits and those that have been detected have generally been attributed to circumstellar interaction.
Assuming a standard relation between X-ray luminosity $L_x$ and $\dot E$, Perna et al. 
find that the limits rule out most pulsars being born with periods in the ms range.
On a timescale of $4-10$ yr after the explosion, the heating and ionization of the supernova material by the energetic radiation from a pulsar nebula can lead to observable effects at optical wavelengths \cite{CF92}.
The characteristic increasing velocity with age expected in this situation has not yet been observed.

Observations of neutron stars have shown evidence for a broad range of initial magnetic fields.
Initial periods are more uncertain, but the evidence points to relatively long ($\ge 100$ ms) periods being typical.
The result is that pulsars do not generally manifest themselves in supernovae, but there should be rare cases of initially energetic pulsars that have an observational effect.
We may have already observed such cases, but the evidence for any particular event is not yet secure.


\begin{theacknowledgments}
  I am grateful to the organizers for a pleasant and stimulating meeting.
This research was supported in part by NASA grant NNX07AG78G.
\end{theacknowledgments}



\bibliographystyle{aipproc}   

\bibliography{chevalier}

\hyphenation{Post-Script Sprin-ger}
\begin{thebibliography}{45}
\expandafter\ifx\csname natexlab\endcsname\relax\def\natexlab#1{#1}\fi
\providecommand{\enquote}[1]{``#1''}
\expandafter\ifx\csname url\endcsname\relax
  \def\url#1{\texttt{#1}}\fi
\expandafter\ifx\csname urlprefix\endcsname\relax\def\urlprefix{URL }\fi
\providecommand{\eprint}[2][]{\url{#2}}

\bibitem[{Faucher-Gigu{\`e}re} and {Kaspi}(2006)]{fkaspi06}
C.~{Faucher-Gigu{\`e}re}, and V.~M. {Kaspi}, \emph{ApJ} \textbf{643}, 332--355
  (2006), \eprint{arXiv:astro-ph/0512585}.

\bibitem[{Popov} et~al.(2010)]{popov10}
S.~B. {Popov}, J.~A. {Pons}, J.~A. {Miralles}, P.~A. {Boldin}, and
  B.~{Posselt}, \emph{MNRAS} \textbf{401}, 2675--2686 (2010),
  \eprint{0910.2190}.

\bibitem[{Chevalier}(2005)]{chev05}
R.~A. {Chevalier}, \emph{ApJ} \textbf{619}, 839--855 (2005),
  \eprint{arXiv:astro-ph/0409013}.

\bibitem[{Gupta} et~al.(2005)]{gupta05}
Y.~{Gupta}, D.~{Mitra}, D.~A. {Green}, and A.~{Acharyya}, \emph{Current
  Science} \textbf{89}, 853--+ (2005), \eprint{arXiv:astro-ph/0508257}.

\bibitem[{Camilo} et~al.(2006)]{camilo06}
F.~{Camilo}, S.~M. {Ransom}, B.~M. {Gaensler}, P.~O. {Slane}, D.~R. {Lorimer},
  J.~{Reynolds}, R.~N. {Manchester}, and S.~S. {Murray}, \emph{ApJ}
  \textbf{637}, 456--465 (2006), \eprint{arXiv:astro-ph/0509823}.

\bibitem[{Halpern} and {Gotthelf}(2010{\natexlab{a}})]{HG10b}
J.~P. {Halpern}, and E.~V. {Gotthelf}, \emph{ArXiv e-prints}
  (2010{\natexlab{a}}), \eprint{1008.2558}.

\bibitem[{Gelfand} and {Gaensler}(2007)]{GG07}
J.~D. {Gelfand}, and B.~M. {Gaensler}, \emph{ApJ} \textbf{667}, 1111--1118
  (2007), \eprint{0706.1054}.

\bibitem[{Gaensler} et~al.(2008)]{gaens08}
B.~M. {Gaensler}, A.~{Tanna}, P.~O. {Slane}, C.~L. {Brogan}, J.~D. {Gelfand},
  N.~M. {McClure-Griffiths}, F.~{Camilo}, C.~{Ng}, and J.~M. {Miller},
  \emph{ApJ} \textbf{680}, L37--L40 (2008), \eprint{0804.0462}.

\bibitem[{Matheson} and {Safi-Harb}(2005)]{math05}
H.~{Matheson}, and S.~{Safi-Harb}, \emph{AdSpR} \textbf{35}, 1099--1105 (2005),
  \eprint{arXiv:astro-ph/0504369}.

\bibitem[{Halpern} and {Gotthelf}(2010{\natexlab{b}})]{HG10a}
J.~P. {Halpern}, and E.~V. {Gotthelf}, \emph{ApJ} \textbf{709}, 436--446
  (2010{\natexlab{b}}), \eprint{0911.0093}.

\bibitem[{Gotthelf} and {Halpern}(2009)]{GH09}
E.~V. {Gotthelf}, and J.~P. {Halpern}, \emph{ApJ} \textbf{695}, L35--L39
  (2009), \eprint{0902.3007}.

\bibitem[{Gotthelf} and {Halpern}(2007)]{GH07}
E.~V. {Gotthelf}, and J.~P. {Halpern}, \emph{ApJ} \textbf{664}, L35--L38
  (2007), \eprint{0704.2255}.

\bibitem[{Ho} and {Heinke}(2009)]{HH09}
W.~C.~G. {Ho}, and C.~O. {Heinke}, \emph{Nature} \textbf{462}, 71--73 (2009),
  \eprint{0911.0672}.

\bibitem[{Duncan} and {Thompson}(1992)]{DT92}
R.~C. {Duncan}, and C.~{Thompson}, \emph{ApJ} \textbf{392}, L9--L13 (1992).

\bibitem[{Vink} and {Kuiper}(2006)]{vink06}
J.~{Vink}, and L.~{Kuiper}, \emph{MNRAS} \textbf{370}, L14--L18 (2006),
  \eprint{arXiv:astro-ph/0604187}.

\bibitem[{Rutkowski} et~al.(2010)]{rutkowski10}
M.~J. {Rutkowski}, E.~M. {Schlegel}, J.~W. {Keohane}, and R.~A. {Windhorst},
  \emph{ApJ} \textbf{715}, 908--918 (2010), \eprint{1005.0635}.

\bibitem[{Kaplan} et~al.(2004)]{kaplan04}
D.~L. {Kaplan}, D.~A. {Frail}, B.~M. {Gaensler}, E.~V. {Gotthelf}, S.~R.
  {Kulkarni}, P.~O. {Slane}, and A.~{Nechita}, \emph{ApJS} \textbf{153},
  269--315 (2004), \eprint{arXiv:astro-ph/0403313}.

\bibitem[{Muno} and {et al.}(2006)]{muno06}
M.~P. {Muno}, and {et al.}, \emph{ApJ} \textbf{636}, L41--L44 (2006),
  \eprint{arXiv:astro-ph/0509408}.

\bibitem[{Figer} et~al.(2005)]{figer05}
D.~F. {Figer}, F.~{Najarro}, T.~R. {Geballe}, R.~D. {Blum}, and R.~P.
  {Kudritzki}, \emph{ApJ} \textbf{622}, L49--L52 (2005),
  \eprint{arXiv:astro-ph/0501560}.

\bibitem[{Davies} et~al.(2009)]{davies09}
B.~{Davies}, D.~F. {Figer}, R.~{Kudritzki}, C.~{Trombley}, C.~{Kouveliotou},
  and S.~{Wachter}, \emph{ApJ} \textbf{707}, 844--851 (2009),
  \eprint{0910.4859}.

\bibitem[{Vasisht} and {Gotthelf}(1997)]{VG97}
G.~{Vasisht}, and E.~V. {Gotthelf}, \emph{ApJ} \textbf{486}, L129+ (1997),
  \eprint{arXiv:astro-ph/9706058}.

\bibitem[{Serafimovich} et~al.(2005)]{sera05}
N.~I. {Serafimovich}, P.~{Lundqvist}, Y.~A. {Shibanov}, and J.~{Sollerman},
  \emph{AdSpR} \textbf{35}, 1106--1111 (2005), \eprint{arXiv:astro-ph/0501523}.

\bibitem[{Krause} et~al.(2008)]{krause08}
O.~{Krause}, S.~M. {Birkmann}, T.~{Usuda}, T.~{Hattori}, M.~{Goto}, G.~H.
  {Rieke}, and K.~A. {Misselt}, \emph{Science} \textbf{320}, 1195-- (2008),
  \eprint{0805.4557}.

\bibitem[{Young} et~al.(2006)]{young06}
P.~A. {Young}, C.~L. {Fryer}, A.~{Hungerford}, D.~{Arnett}, G.~{Rockefeller},
  F.~X. {Timmes}, B.~{Voit}, C.~{Meakin}, and K.~A. {Eriksen}, \emph{ApJ}
  \textbf{640}, 891--900 (2006), \eprint{arXiv:astro-ph/0511806}.

\bibitem[{Arnett} et~al.(1989)]{arnett89}
W.~D. {Arnett}, J.~N. {Bahcall}, R.~P. {Kirshner}, and S.~E. {Woosley},
  \emph{ARA\&A} \textbf{27}, 629--700 (1989).

\bibitem[{Bodenheimer} and {Ostriker}(1974)]{bo74}
P.~{Bodenheimer}, and J.~P. {Ostriker}, \emph{ApJ} \textbf{191}, 465--472
  (1974).

\bibitem[{Gaffet}(1977{\natexlab{a}})]{gaffet77a}
B.~{Gaffet}, \emph{ApJ} \textbf{216}, 565--577 (1977{\natexlab{a}}).

\bibitem[{Gaffet}(1977{\natexlab{b}})]{gaffet77b}
B.~{Gaffet}, \emph{ApJ} \textbf{216}, 852--864 (1977{\natexlab{b}}).

\bibitem[{Maeda} et~al.(2007)]{maeda07}
K.~{Maeda}, M.~{Tanaka}, K.~{Nomoto}, N.~{Tominaga}, K.~{Kawabata}, P.~A.
  {Mazzali}, H.~{Umeda}, T.~{Suzuki}, and T.~{Hattori}, \emph{ApJ}
  \textbf{666}, 1069--1082 (2007), \eprint{0705.2713}.

\bibitem[{Kasen} and {Bildsten}(2010)]{kasen10}
D.~{Kasen}, and L.~{Bildsten}, \emph{ApJ} \textbf{717}, 245--249 (2010),
  \eprint{0911.0680}.

\bibitem[{Woosley}(2010)]{woosley10}
S.~E. {Woosley}, \emph{ApJ} \textbf{719}, L204--L207 (2010),
  \eprint{0911.0698}.

\bibitem[{Chevalier}(1989)]{chev89}
R.~A. {Chevalier}, \emph{ApJ} \textbf{346}, 847--859 (1989).

\bibitem[{Komissarov} and {Barkov}(2007)]{kom07}
S.~S. {Komissarov}, and M.~V. {Barkov}, \emph{MNRAS} \textbf{382}, 1029--1040
  (2007), \eprint{0707.0264}.

\bibitem[{Bucciantini} et~al.(2008)]{bucc08}
N.~{Bucciantini}, E.~{Quataert}, J.~{Arons}, B.~D. {Metzger}, and T.~A.
  {Thompson}, \emph{MNRAS} \textbf{383}, L25--L29 (2008), \eprint{0707.2100}.

\bibitem[{Miller} and {et al.}(2009)]{miller09}
A.~A. {Miller}, and {et al.}, \emph{ApJ} \textbf{690}, 1303--1312 (2009),
  \eprint{0808.2193}.

\bibitem[{Gezari} and {et al.}(2009)]{gezari09}
S.~{Gezari}, and {et al.}, \emph{ApJ} \textbf{690}, 1313--1321 (2009),
  \eprint{0808.2812}.

\bibitem[{Pastorello} and {et al.}(2010)]{past10}
A.~{Pastorello}, and {et al.}, \emph{ArXiv e-prints}  (2010),
  \eprint{1008.2674}.

\bibitem[{Chevalier}(1977)]{chev77}
R.~A. {Chevalier}, \enquote{{Was SN 1054 A Type II Supernova?},} in
  \emph{Supernovae}, edited by {D.~N.~Schramm}, 1977, vol.~66 of
  \emph{Astrophysics and Space Science Library}, pp. 53--+.

\bibitem[{Gal-Yam} and {et al.}(2009)]{galyam09}
A.~{Gal-Yam}, and {et al.}, \emph{Nature} \textbf{462}, 624--627 (2009),
  \eprint{1001.1156}.

\bibitem[{Patat} and {et al.}(2001)]{patat01}
F.~{Patat}, and {et al.}, \emph{ApJ} \textbf{555}, 900--917 (2001),
  \eprint{arXiv:astro-ph/0103111}.

\bibitem[{Graves} and {et al.}(2005)]{graves05}
G.~J.~M. {Graves}, and {et al.}, \emph{ApJ} \textbf{629}, 944--959 (2005),
  \eprint{arXiv:astro-ph/0505066}.

\bibitem[{Ng} et~al.(2009)]{ng09}
C.~{Ng}, B.~M. {Gaensler}, S.~S. {Murray}, P.~O. {Slane}, S.~{Park},
  L.~{Staveley-Smith}, R.~N. {Manchester}, and D.~N. {Burrows}, \emph{ApJ}
  \textbf{706}, L100--L105 (2009), \eprint{0910.3610}.

\bibitem[{Bietenholz} et~al.(2010)]{bieten10}
M.~F. {Bietenholz}, N.~{Bartel}, and M.~P. {Rupen}, \emph{ApJ} \textbf{712},
  1057--1069 (2010).

\bibitem[{Perna} et~al.(2008)]{perna08}
R.~{Perna}, R.~{Soria}, D.~{Pooley}, and L.~{Stella}, \emph{MNRAS}
  \textbf{384}, 1638--1648 (2008), \eprint{0712.1040}.

\bibitem[{Chevalier} and {Fransson}(1992)]{CF92}
R.~A. {Chevalier}, and C.~{Fransson}, \emph{ApJ} \textbf{395}, 540--552 (1992).

\end{thebibliography}

\IfFileExists{\jobname.bbl}{}
 {\typeout{}
  \typeout{******************************************}
  \typeout{** Please run "bibtex \jobname" to optain}
  \typeout{** the bibliography and then re-run LaTeX}
  \typeout{** twice to fix the references!}
  \typeout{******************************************}
  \typeout{}
 }

\end{document}